\documentclass[useAMS,usenatbib]{mn2e}
\usepackage[dvips]{graphicx,color}
\usepackage{amssymb}
\usepackage{indentfirst}
\usepackage{epsfig}
\usepackage{pdflscape}
\usepackage{afterpage}
\usepackage{multicol}

\newcommand{\Ka}{\ensuremath{\hbox{K}\alpha~}}

\newcommand{\Kb}{\ensuremath{\hbox{K}\beta~}}

\def\nicer{{\it NICER }}
\def\nus{{\it NuSTAR }}
\def\fermi{{\it Fermi}}
\def\swift{{\it Swift}}
\def\hxmt{{\it HXMT }}

\begin{document}
\title[]{Disk vs wind accretion in X-ray pulsar GX 301-2}

\author[J. Liu et al.]{Jiren Liu$^{1}$\thanks{E-mail: liujiren@bjp.org.cn}
, Long Ji$^{2,7}$, Peter A. Jenke$^3$, Victor Doroshenko$^2$, Zhenxuan Liao$^4$, 
\newauthor
Xiaobo Li$^5$, Shuangnan Zhang$^5$, Mauro Orlandini$^6$, Mingyu Ge$^5$, Shu Zhang$^5$, 
\newauthor
Andrea Santangelo$^2$\\
$^{1}$Beijing Planetarium, Xizhimenwai Road, Beijing 100044, China\\
$^{2}$Institut f\"ur Astronomie und Astrophysik, Kepler Center for Astro and Particle Physics, Eberhard Karls Universit\"at, Sand 1,\\ 72076 T\"ubingen, Germany\\
$^{3}$University of Alabama in Huntsville, Huntsville, AL 35812, USA\\
$^{4}$National Astronomical Observatories, 20A Datun Road, Beijing 100012, China\\
$^{5}$ Key Laboratory for Particle Astrophysics, Institute of High Energy Physics, Beijing 100049, China\\
$^{6}$ INAF Osservatorio di Astrofisica e Scienza dello Spazio di Bologna, Via Piero Gobetti 101, I-40129 Bologna, Italy\\
$^{7}$ School of physics and astronomy, Sun Yat-Sen University, Zhuhai, Guangdong 519082, China \\
}

\date{}

\maketitle

\begin{abstract}
GX 301-2 provides
a rare opportunity to study both disk and wind accretion in a same target.
We report {\it Insight}-\hxmt observations of the spin-up event of GX 301-2 
happened in 2019 and compare with those of wind-fed state.
The pulse profiles of the 
initial rapid spin-up period are dominated by one main peak, while those of the later 
slow spin-up period are composed of two similar peaks, as those of
wind-fed state. These behaviors are confirmed by \fermi/GBM data, which also show 
that during the rapid spin-up period, the main peak increases with luminosity up to
$8\times10^{37}$ erg\,s$^{-1}$, but the faint peak keeps almost constant. 
The absorption column densities during the spin-up period are 
$\sim1.5\times10^{23}$\,cm$^{-2}$, much less than those of wind-fed state at similar
luminosity ($\sim9\times10^{23}$\,cm$^{-2}$), supporting the scenario
that most of material is condensed into a disk during the spin-up period.
We discuss possible differences between disk and wind accretion 
 that may explain the observed different trend of pulse profiles.

\end{abstract}


\section{Introduction}

X-ray pulsars are powered by accretion of material from a normal companion star 
onto a rotating magnetized neutron star. 
The mass transfer process could be through either 
a stellar wind or a Roche-Lobe overflow (RLOF).
For RLOF, the transferred material has sufficient angular momentum to form 
a rotating disk, through which the matter spirals towards the neutron star \citep{PR72}.
On the other hand, the transfer of stellar wind resembles  Bondi-Hoyle
accretion with a quasi-spherical distribution \citep{DO73}, 
although formation of short-lived transient accretion disks is also possible 
\citep{FT88, BP09, XS19}.
The disk accretion transfers angular momentum to the neutron star and
leads to variations of 
spin frequency of the neutron star measurable on days timescale,
while wind accretion results in stochastic spin-frequency changes on shorter timescales.
Both the disk and wind flow will be dominated by the magnetic field within 
the magnetosphere and will be channeled along the magnetic field lines to the 
surface of the neutron star \citep[e.g.][]{Dav73,BS76}. 

Generally, mass transfer in X-ray pulsars is dominated by one of mechanisms,
 either wind or disk accretion, but there are 
a few X-ray pulsars showing both wind and disk accretion alternately.
These sources are ideal targets to study 
the differences between different accretion processes, since the major parameters 
of the system, such as the magnetic field and inclination angle, are not changed.
The classical wind-fed X-ray pulsar, GX 301-2, is such a source. 

GX 301-2 is a slow rotator with a pulse period $\sim680$ s \citep{Whi76},
an orbital period $\sim41.5$ days and an orbital
eccentricity $\sim0.46$ \citep{Sat86,Koh97}. 
Its optical companion, Wray 977, has a mass of $39-53 M_\odot$ and 
a radius of $62 R_\odot$, close to the Roche lobe radius near periastron \citep{Kap06}.
Its spin history monitored by BATSE on {\it Compton} observatory and GBM on \fermi\  
spacecraft is composed of long period of small frequency variability and 
several rapid spin-up events lasting for tens of days \citep{Koh97}.
The long periods of small stochastic spin frequency changes indicate wind accretion, 
while the spin-up events implies transient disk accretion.
Indeed, during the spin-up events, its spin derivatives were found to be correlated with 
fluxes ($\dot{\nu}\propto F^{0.75\pm0.05}$), consistent with a scenario of 
disk accretion \citep{Liu20}. There may be also some correlations between fluxes and 
spin derivatives during wind state \citep{Dor10}.

A strong spin-up event of GX 301-2 happened between Dec. 2018 and Mar. 2019.
\citet{Nab19} reported \nus observations of GX 301-2 
near the end of this spin-up event (Mar. 3, 2019, MJD 58545) and found no 
significant differences in spectral 
and temporal properties compared with normal wind-fed state. In contrast, 
with X-Calibur polarimeter, \nicer, and \swift\  observations of GX 301-2 during 
the initial spin-up period (around Dec. 31, 2018, MJD 58483), 
\citet{Aba20} found that its pulse profiles are strongly dominated by one main peak, 
quite different from two approximately equal peaks in normal wind-fed state.
In this paper we analyze Insight Hard X-ray Modulation Telescope ({\it HXMT}) 
observations 
of GX 301-2 during the spin-up episode performed around Jan. 11-12, 2019,
i.e. right in the middle of this event with the 
aim to study the differences between wind and disk accretion of GX 301-2.

\begin{figure}
	\hspace{-0.0cm}
	\includegraphics[width=3.4in]{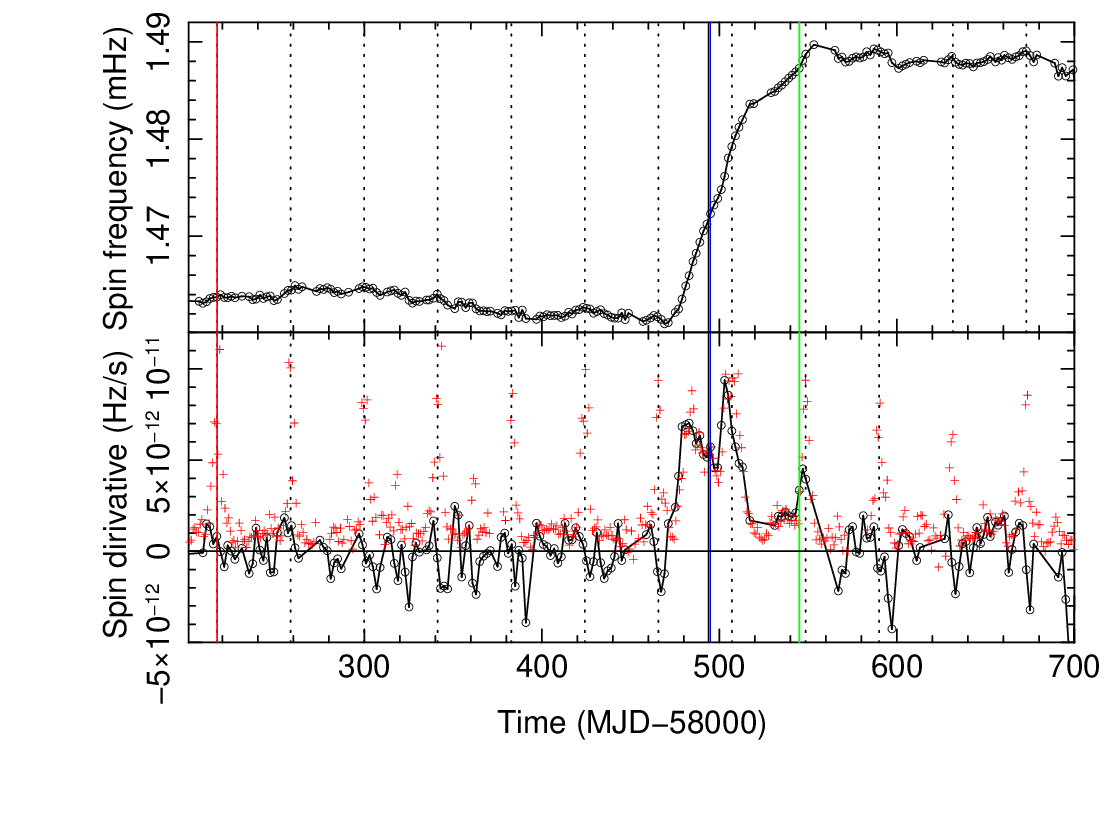}
	\caption{The 2019 spin-up event of GX 301-2 monitored by \fermi/GBM (top panel), 
	together with the spin derivatives (circles in bottom panel).
The corresponding \swift/BAT 15-50 keV fluxes (multiplied by a factor of 
$5\times10^{-11}$, red pluses) are over-plotted.
Vertical dotted lines indicate the time of periastron, and vertical solid lines 
indicate dates of \hxmt observations analyzed.
}
\end{figure}

\begin{figure*}
	\includegraphics[width=3.3in]{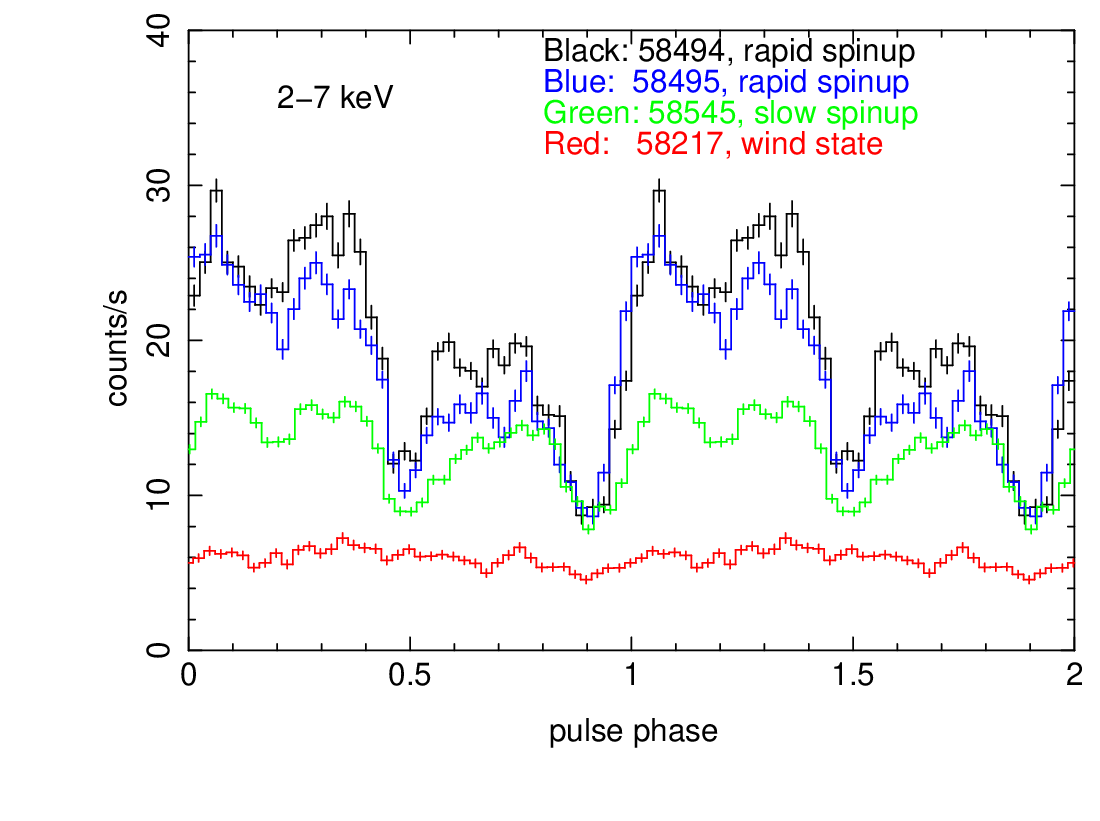}
	\includegraphics[width=3.3in]{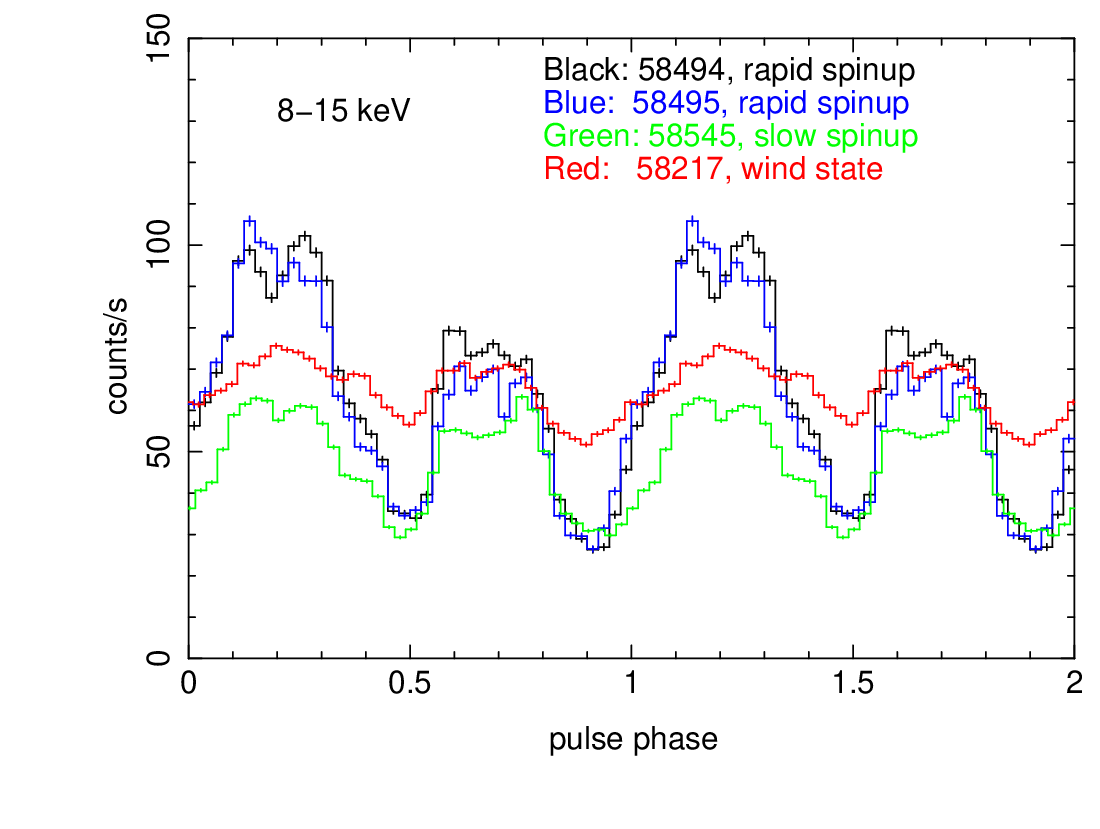}
	\includegraphics[width=3.3in]{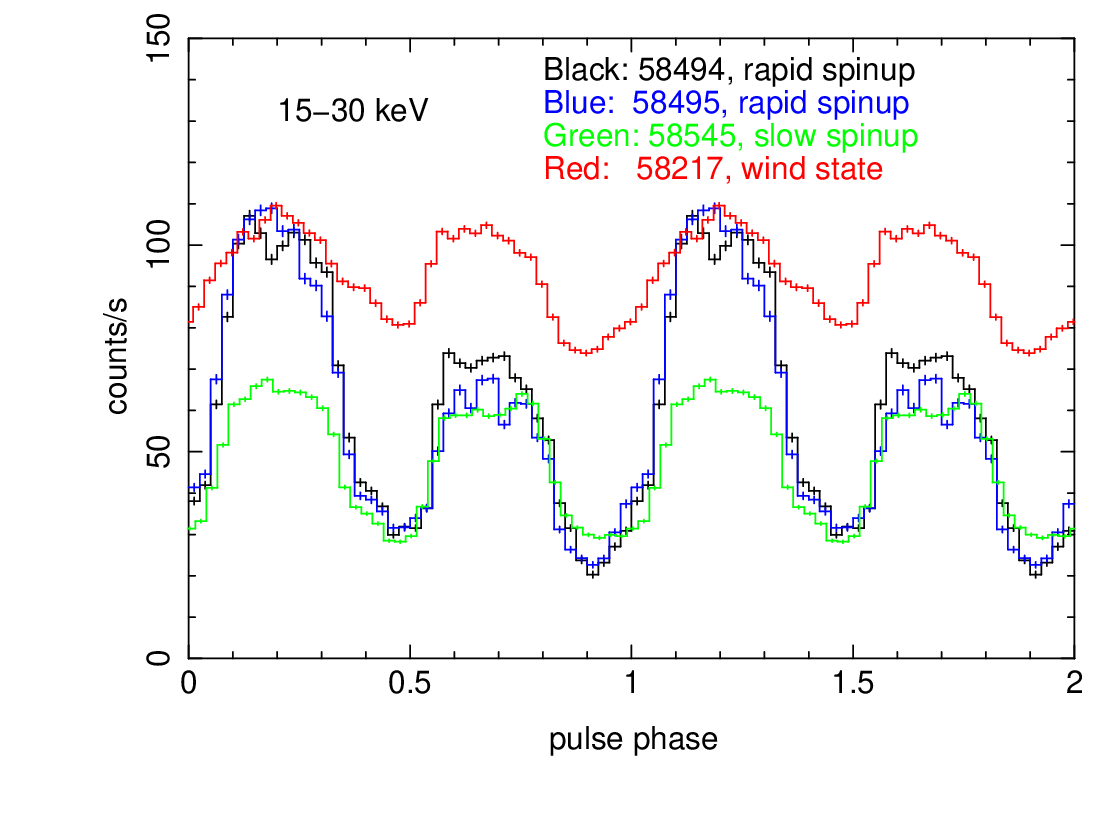}
	\includegraphics[width=3.3in]{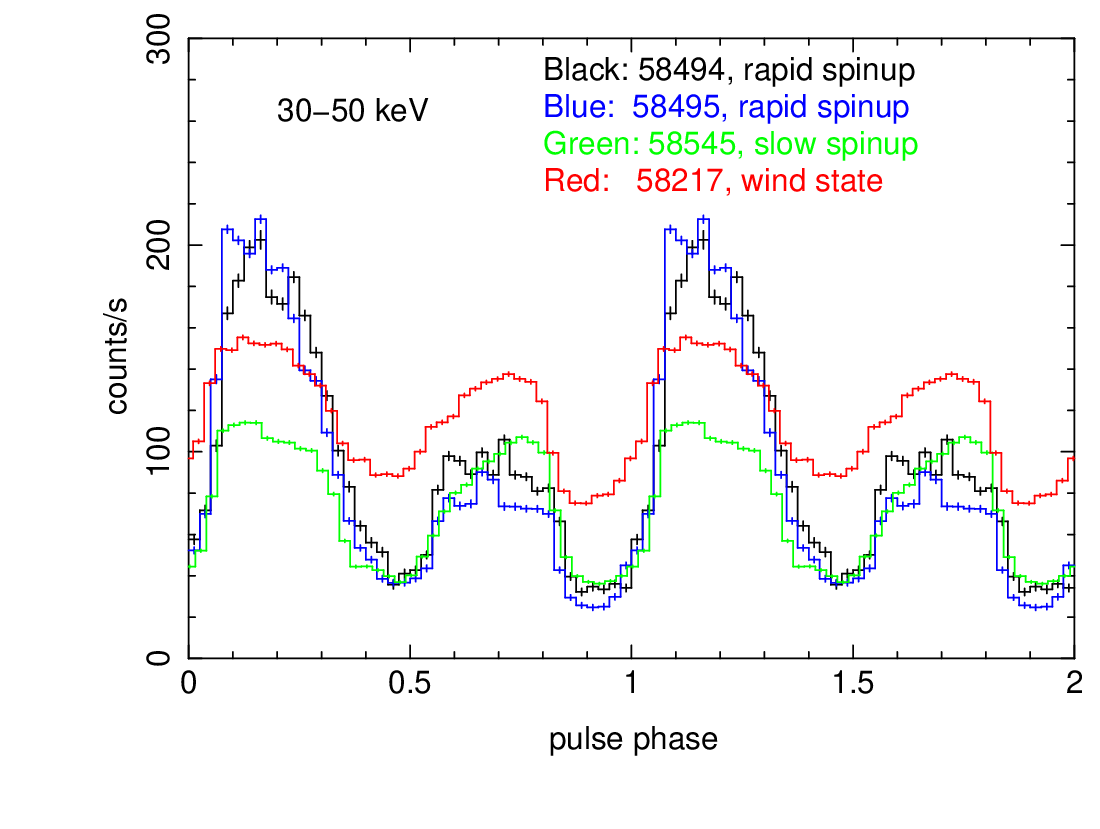}
	\caption{Pulse profiles of GX 301-2 observed by \hxmt around 
	MJD 58494, 59495, 58545 and 58217 at four different energy bands.
The profiles of the rapid spin-up period (MJD 58494 and 59495) are 
dominated by one main peak, while others are composed of two similar peaks.
MJD 58217 represents a typical wind-fed state with a flux similar to those 
of the rapid spin-up period (MJD 58494 and 59495).
}
\end{figure*}

\begin{figure*}
	\includegraphics[width=3.2in]{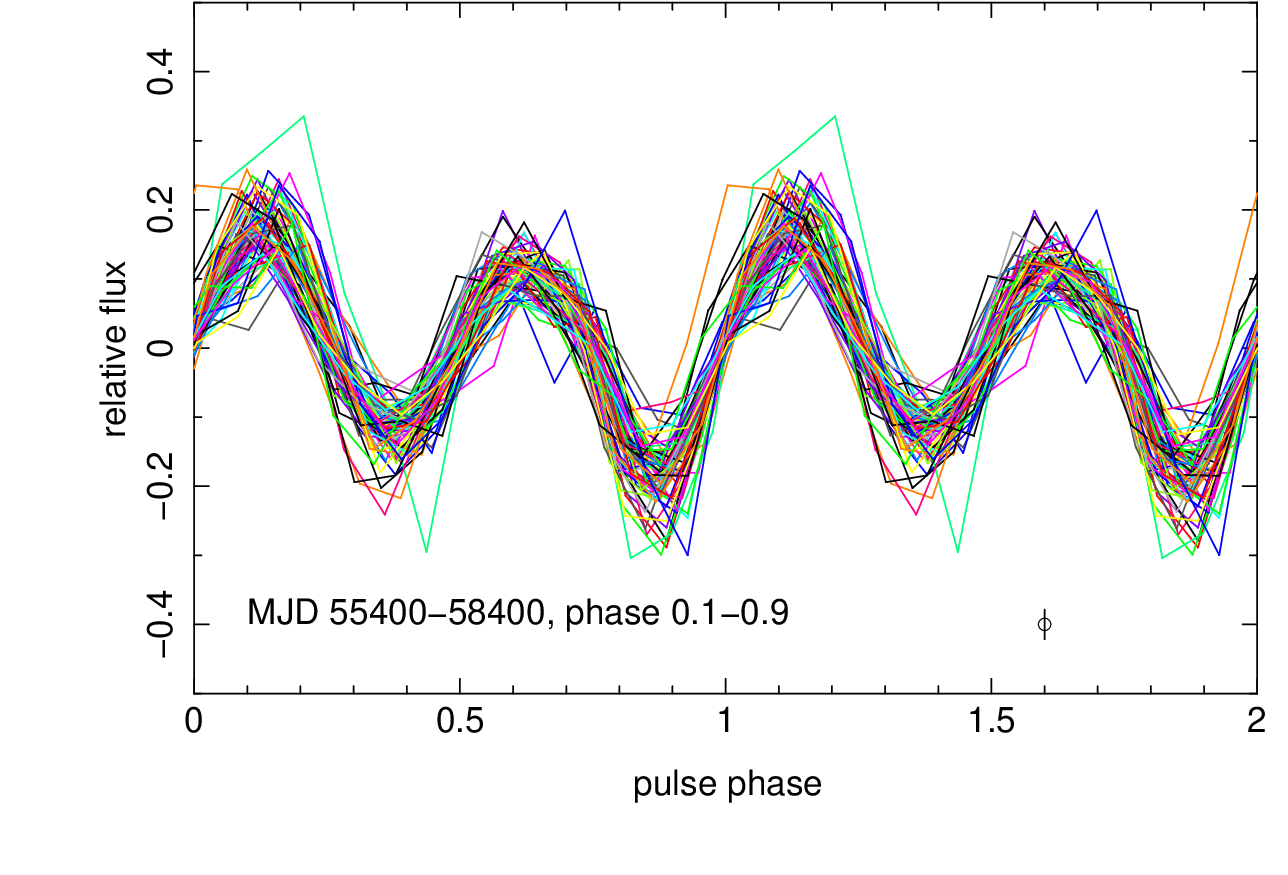}
	\includegraphics[width=3.2in]{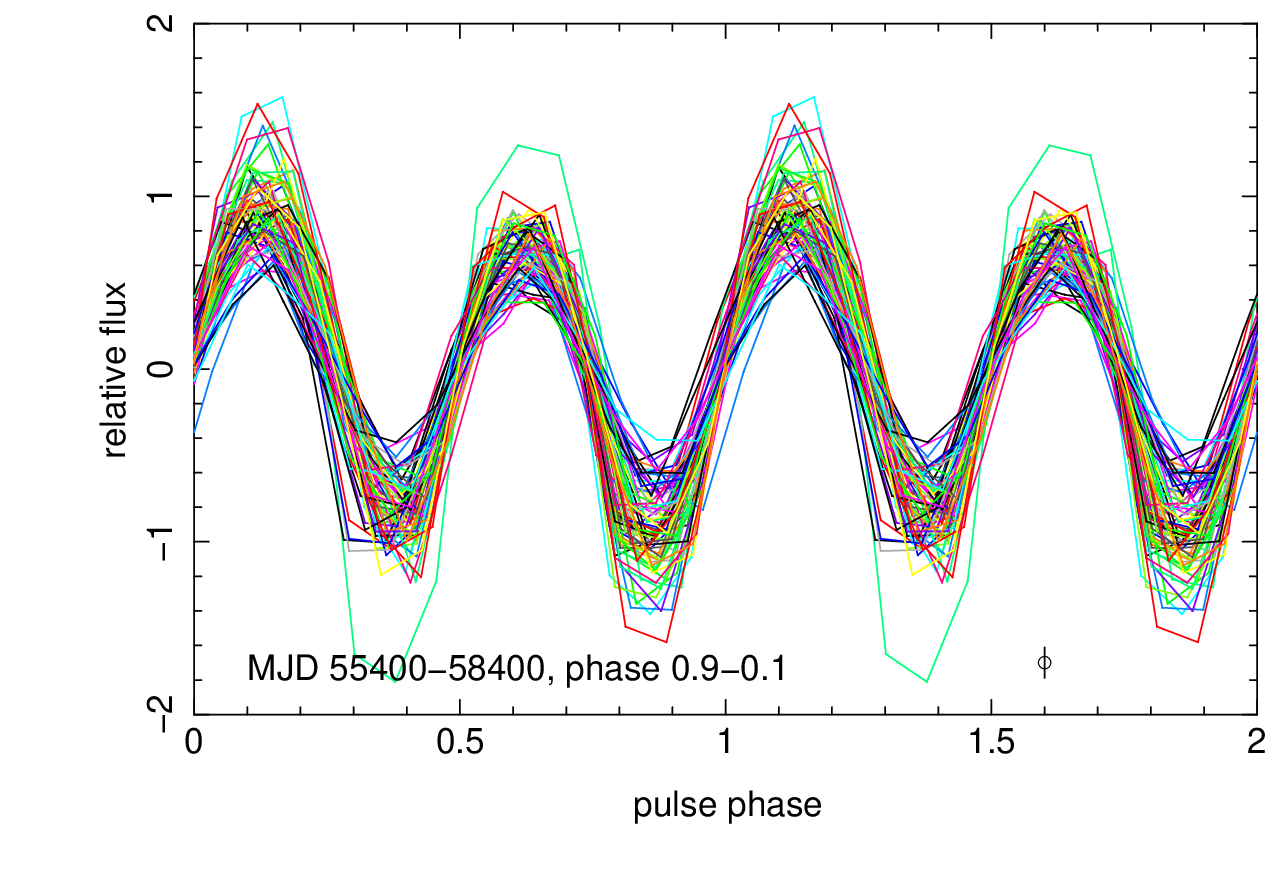}
	\includegraphics[width=3.2in]{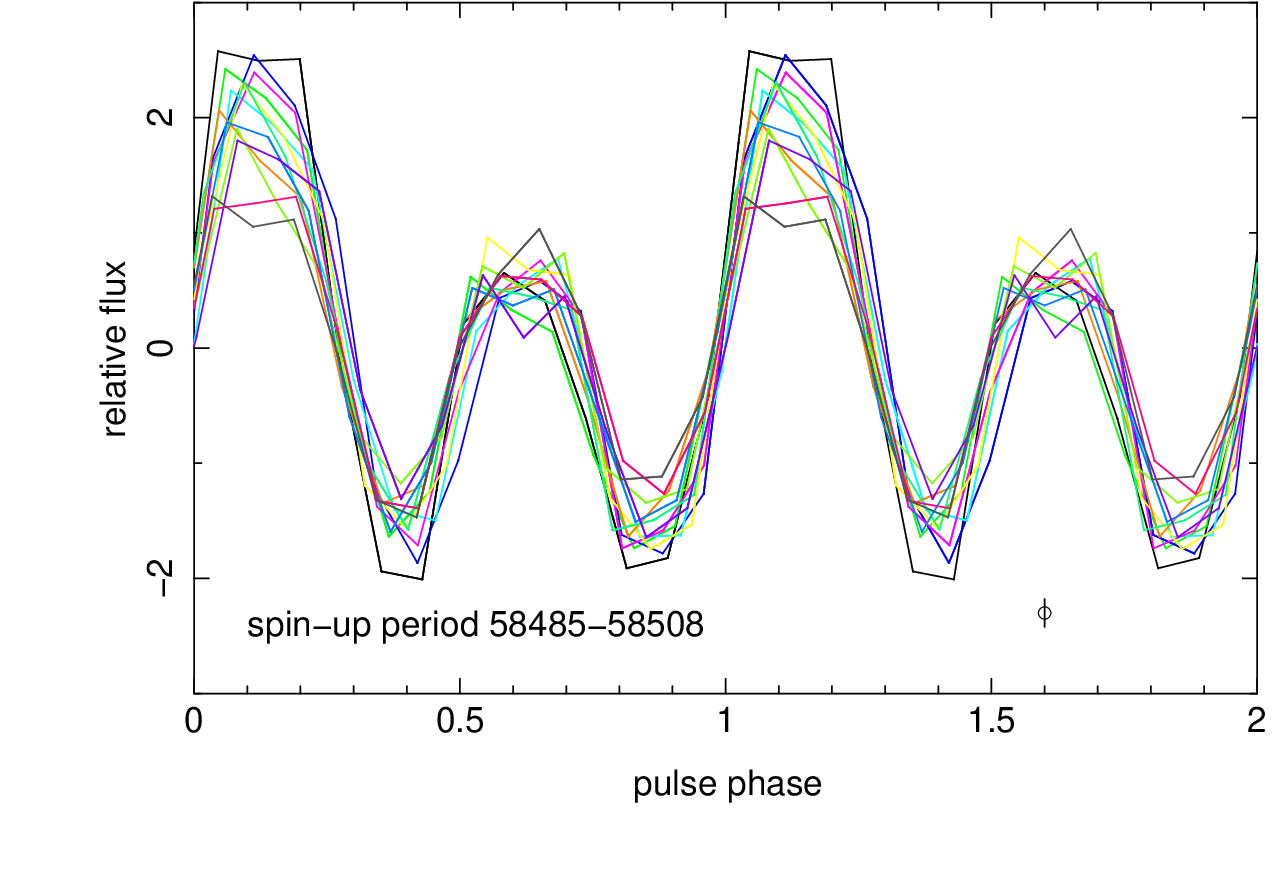}
	\includegraphics[width=3.2in]{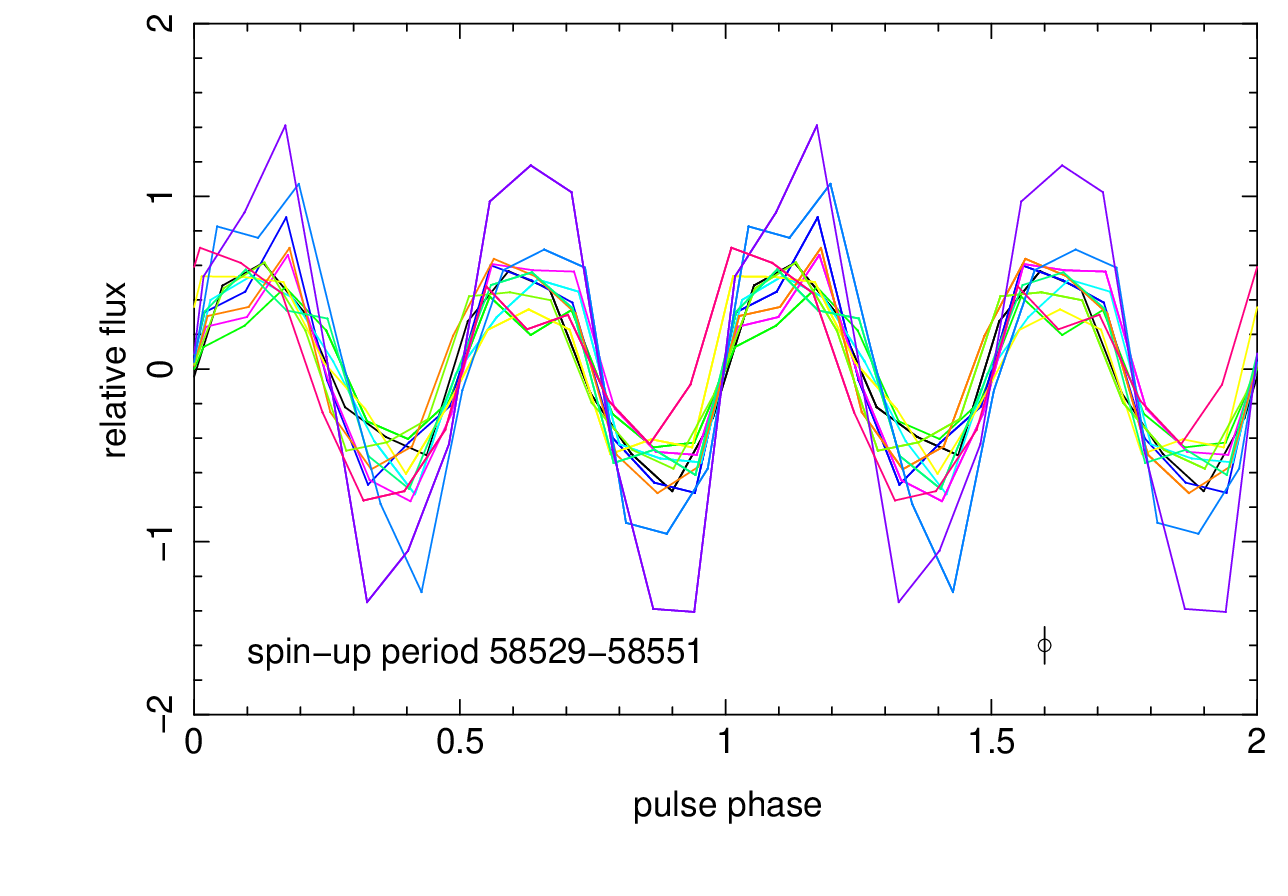}
	\caption{12-25 keV pulse profile of GX 301-2 extracted from \fermi/GBM data 
	during non spin-up periods and 2019 spin-up episode.
	The non spin-up periods are divided into orbital phase intervals
	of 0.1-0.9 and 0.9-0.1, while the spin-up episode is divided into 
	initial rapid spin-up and later slow spin-up phase. The inset circles 
represent mean errors of the relative fluxes. Different colors correspond
to measurements at different intervals.
}
\end{figure*}

\begin{figure}
	\includegraphics[width=3.3in]{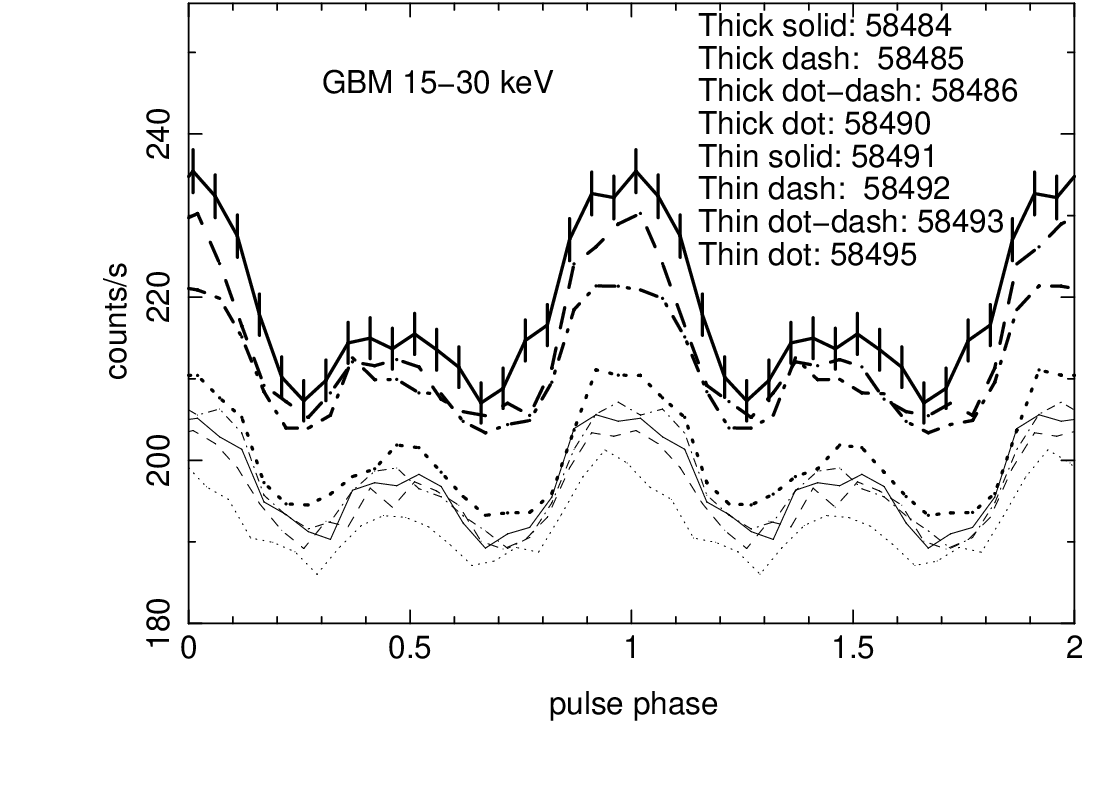}
	\caption{15-30 keV pulse profile of GX 301-2 extracted from \fermi/GBM 
	data based on individual daily extractions. For clarity, only the errors of 
	MJD 58484 are plotted.
}
\end{figure}

\begin{figure}
	\includegraphics[width=3.3in]{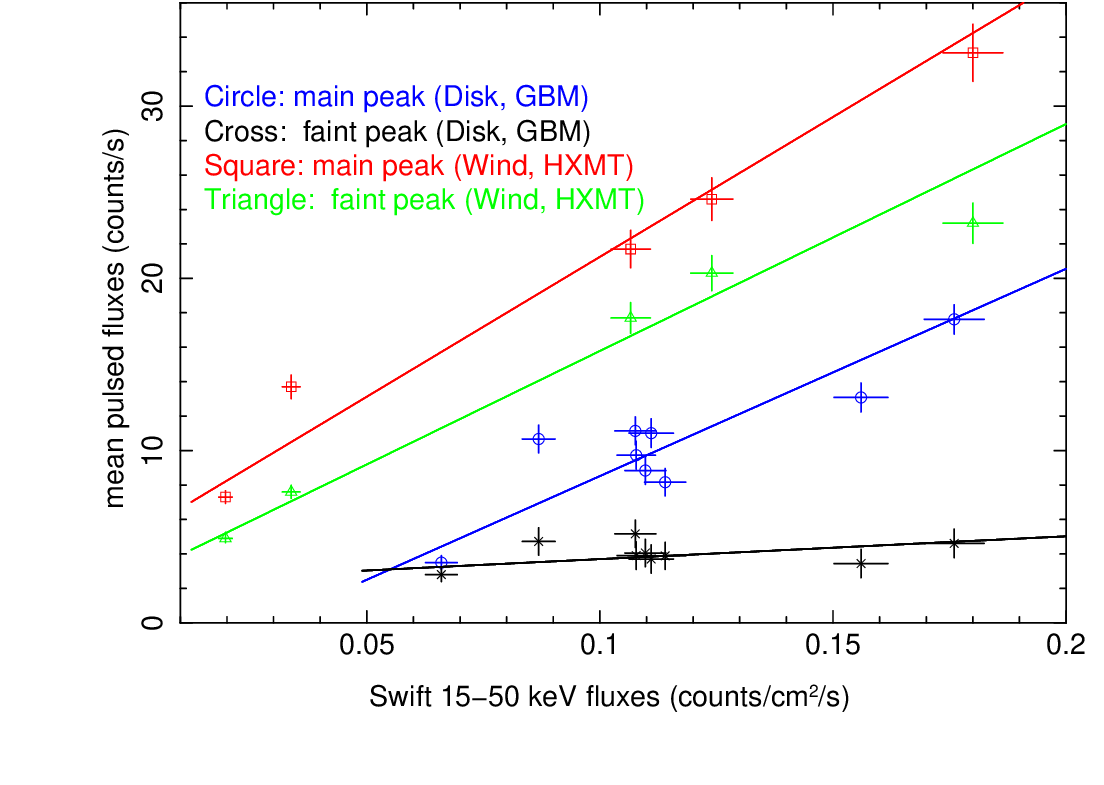}
	\caption{15-30 keV mean pulsed fluxes of the main and faint peaks
		plotted against the \swift\  daily 15-50 keV fluxes for disk 
		(extracted from Fermi/GBM) and wind (extracted from {\it HXMT}) state. 
		   The solid lines are linear fits to the data. 
The two data points of disk state around 0.06-0.07 are the \hxmt results on
MJD 58545 (SD) extrapolated to \fermi/GBM values.
}
\end{figure}

\section{Observation data}

{\it Insight}-\hxmt is a Chinese X-ray satellite launched on June, 2017. It carries 
three collimated instruments sensitive to different energy bands: low energy telescope
(LE, 1-15 keV), medium energy telescope (ME, 5-30 keV), and high energy
telescope (HE, 20-250 keV). 
The corresponding effective areas are 384, 952, and 5100 cm$^2$, respectively. 
For details of \hxmt we refer to \citet{Zhang20} and references 
therein.
The Gamma-ray Burst Monitor \citep[GBM,][]{GBM09} on the \fermi\  spacecraft 
is continuously monitoring the spin histories of X-ray
pulsars\footnote{https://gammaray.msfc.nasa.gov/gbm/science/pulsars/} \citep{Fin09,Mal20}. 

\fermi/GBM detected one strong spin-up event of GX 301-2 around Jan., 2019, 
which is presented in Figure 1. 
The spin-up event started around an orbital phase 0.22 (Dec. 23, 2018, MJD 58475)
and reached a rapid spin-up rate $\sim6\times10^{-12}$ Hz/s.
The spin-up rate was further increased a little bit
($\sim1\times10^{-11}$ Hz/s) just prior to periastron. After periastron, 
the spin-up rate slowed down to 
about $2\times10^{-12}$ Hz/s, and then increased to $\sim4\times10^{-12}$ Hz/s
near the second periastron. After the second periastron, the spin-up 
event ended. 
Good correspondence between 
the spin-up rates and the \swift/BAT fluxes can be seen during the spin-up episode
\citep[see][for more details]{Liu20}.

\hxmt observed GX 301-2 twice during the initial rapid spin-up period
on Jan. 11 and 12, 2019 (MJD 58494 and 58495), and once during the later slow 
spin-up phase (Mar. 3, 2019, MJD 58545). The later observation was taken 
on the same day with \nus discussed by \citet{Nab19}.
These \hxmt observations are complementary to X-Calibur, \swift/XRT, and 
\nicer observations reported previously, since \hxmt provides simultaneous 
broad band data during the middle of the rapid spin-up period.
For comparison, we also analyzed other \hxmt observations of GX 301-2 before 
the 2019 spin-up event. The observation details are listed in Table 1.
The data are processed using HXMTDAS2.03 following the analysis guide.
Both barycentric and binary orbital corrections are applied to the light curves, 
and the ephemeris parameters from \citet{Dor10} are adopted.

\begin{table}
   \caption{List of \hxmt observations}
   \tabcolsep 2.8pt
   \begin{tabular}{ccccc}
      \hline
		 ObsID & $t_{\rm exp}$ (ks)& Obs date & MJD & note\\
      \hline
		 P0101309019 &  1.6 & 2019-01-11&58494 & rapid spin-up (D)\\
		 P0101309020 &  2.9 & 2019-01-12&58495 & rapid spin-up (D)\\
		 P0101309021 &  14.2 & 2019-03-03&58545& slow spin-up (SD)\\
		 P0101309015 &  10.2 & 2018-04-09&58217& wind state (W)\\
      \hline
   \end{tabular}
   \begin{description}
         \begin{footnotesize}
         \item
            Note: $t_{\rm exp}$ refers to the effective exposure of the ME telescope.
         \end{footnotesize}
   \end{description}
\end{table}

\section{Timing analysis}

To study the energy dependences of the pulse profile, we adopted four energy 
bands: 2-7, 8-15, 15-30, and 30-50 keV. The photons within 2-7 keV are from 
LE detectors, 8-30 keV from ME detectors, and 30-50 keV from HE detectors.
We extracted the pulse profiles of different energies from initial 
rapid spin-up periods, MJD 58494 and 58495, which we refer as "D" (disk state).
To compare with those observed during the wind-fed state, we also extracted 
the pulse profiles 
from the observation taken on MJD 58217, when the source was observed 
during the so-called pre-periastron flare at flux level comparable 
to that during the rapid spin-up episode. At this time the spin evolution observed 
by GBM indicates no apparent spin-up trend and we refer as "W" (wind state).  
We also extracted the profiles from the observation on MJD 58545, which is 
referred as "SD" (slow spin-up state).
The pulse periods are calculated from the spin measurement by \fermi/GBM. 
While the phase of MJD 58494 and 58495 are calculated assuming a constant 
spin-up rate, those of MJD 58217 and 58545 are shifted to match the profiles of 
MJD 58494 and 58495.
The extracted pulse profiles for all four energy bands and four observations 
are presented in Figure 2.

As can be seen, most of the profiles within one spin period are composed of two peaks,
which are separated by half period.
During the rapid spin-up periods (MJD 58494-58495, D) one of the peaks of the pulse
profile is significantly higher 
than the other one, and we refer the higher peak as main peak (phase -0.08-0.42) 
and the fainter one 
as faint peak (phase 0.42-0.92). We define the peak fluence as the pulsed flux integrated
over time, and the pulsed flux means the flux minus the minimum flux during the 
whole spin period. 
The fluence ratio between the main and faint peaks ($f_{MF}$) is about 1.6 for photons 
within 8-30 keV for MJD 58494 and is about 2.0 for MJD 58495.
This behavior is similar to X-Calibur result observed around MJD 58483 \citep{Aba20}. 
For 30-50 keV, the main-to-faint contrast is higher, and
$f_{MF}$ is $2.3$ for MJD 58494 and 2.8 for MJD 58495.

In contrast, the profiles of the slow spin-up period (MJD 58545, SD) show two similar 
peaks, similar to the \nus results observed on the same day \citep{Nab19}.
The corresponding $f_{MF}$ of MJD 58545 are about 1.2-1.3 for photons within 8-30 keV
and has a similar value within 30-50 keV.
We note that for photons above 15 keV, the higher fluence ratio of 
MJD 58494 and 58495 (D) are mainly due to the higher fluxes of the main peak.
That is, the fluxes of the main peak of the rapid spin-up periods 
increase significantly compared with those of the slow spin-up period, while
the fluxes of the faint peak do not increase that much.

For the wind-fed state (MJD 58217, W), the two peaks
have similar fluxes for all energies, except for the lowest energies (2-7 keV),
where the pulse profile becomes flat. The corresponding $f_{MF}$ of MJD 58217 are 
about 1.2-1.3 for photons within both 8-30 keV and 30-50 keV, similar to those 
of the slow spin-up period.
The pulse faction (defined as $\frac{f_{max}-f_{min}}{f_{max}+f_{min}}$) 
of the wind-fed state is apparently smaller than those of the 
spin-up periods. While the fluxes of the wind-fed state (MJD 58217) is similar 
to those of MJD 58494 and 58495 (D) for energies within 8-15 keV and 30-50 keV, 
they are higher than those of MJD 58494 and 58495 within 15-30 keV, as shown by
the apparent higher faint peak of the wind state.

Another feature of the pulse profiles observed during the spin-up period is 
that the lower the energy, the broader the main peak width.
The plateau of the 2-7 keV main peak occupies 
almost half spin period, and it seems to be composed of two separate spikes.
There are also two spikes in the main peaks of 
8-15, but is not as apparent as those of the 2-7 keV main peak.

Considering that the effective exposure of \hxmt observations of the spin-up event 
of GX 301-2 lasts only a few ks, and thus only includes a few pulse cycles
and can be biased by stochastical variations, 
we also investigated pulse profile evolution using \fermi/GBM data.
As \fermi/GBM data is dominated by the background, the constant 
component of the light curve is subtracted and only the pulsed 
profile is obtained \citep[e.g.][]{Fin09}. Since the orbital light curves of GX 301-2 
show flares near periastron, for non spin-up periods (MJD 55400-58400, 3000 days
in total) we divided the orbital phase into two intervals of 0.1-0.9 and 0.9-0.1.
There are three continuous data type for GBM: CTIME data with 8 energy channels
and 0.256\,s time bin, CSPEC data with 128 channels and 4.096\,s bin, and CTTE photon
events data. \fermi/GBM made a measurement of pulse profile and spin period 
of GX 301-2 for every $\sim2$ days interval using CTIME data. All
the 12-25 keV profiles within the corresponding periods 
are presented in Figure 3, with different colors indicating measurements 
at different intervals.

As can be seen, the pulse profiles outside of the spin-up periods 
show two similar peaks, although with different pulsed amplitude
for phase intervals of 0.1-0.9 and 0.9-0.1.
The main-to-faint fluence ratios are around 1.2-1.3 estimated from
the pulsed fluxes.
For the 2019 spin-up event, the main peaks during the initial rapid spin-up 
period (MJD 58485-58508) are much higher than the faint peaks, 
similar to the \hxmt results around MJD 58494 and 58495. While those during the 
slow spin-up period (MJD 58529-58551) show two similar
peaks, similar to the \hxmt and \nus results on MJD 58545.

The high fluxes of GX 301-2 during the initial rapid spin-up period also allow us to 
extract background un-subtracted individual pulse profiles of GX 301-2 
from the \fermi/GBM data, and to study the temporal behavior at 
fluxes higher than those of \hxmt observations. We used GBM CSPEC data from N0 detector, 
which has the best viewing angle for GX 301-2 during these time. 
The time periods are picked up by eyes 
to have clear pulsations around 680 s, and the effective area is required to be 
larger than
80\% of the maximum value during that day. The extracted 15-30 keV profiles on date 
Jan. 1, 2, 3, 7, 8, 9, 10, and 12 are plotted in Figure 4. 
As can be seen, the fluence of the main peak decrease with the increasing date,
while the fluence of the faint peak show less changes. 
In Figure 5, we plot the GBM mean pulsed fluxes of the main and faint peaks
vs the daily \swift/BAT 15-50 keV fluxes. It shows clearly that the main peak fluxes 
increase with the \swift\  fluxes, while the faint peak fluxes not.
The fluence ratio $f_{MF}$ is about 4 for the largest fluxes on
MJD 58484. 

To check the behavior of pulse profile of wind-fed state with luminosity,
we also analyzed \hxmt observations of GX 301-2 with good signal and
good pulse detection. These observations are on MJD 57968, 57969, 58121, 58138, 58148,
and 58217. The measured mean pulsed fluxes of the main and faint peaks 
are over-plotted in Figure 5. The uncertainties of these measurements are dominated
by flux variations of GX 301-2 on time scale of hours for 
wind-fed state, which are hard to estimate due to the limited \hxmt effective
exposures. Nevertheless, the 
main-to-faint ratios are generally much stable, and we have assumed \%5
uncertainties of the measured fluxes. 
Both fluxes of the main and faint peaks 
increase with luminosity for wind-fed state, and no significant 
increase of the main-to-faint ratio is found for the observed luminosity.

\begin{figure*}
	\hspace{-0.0cm}
	\includegraphics[width=3.4in]{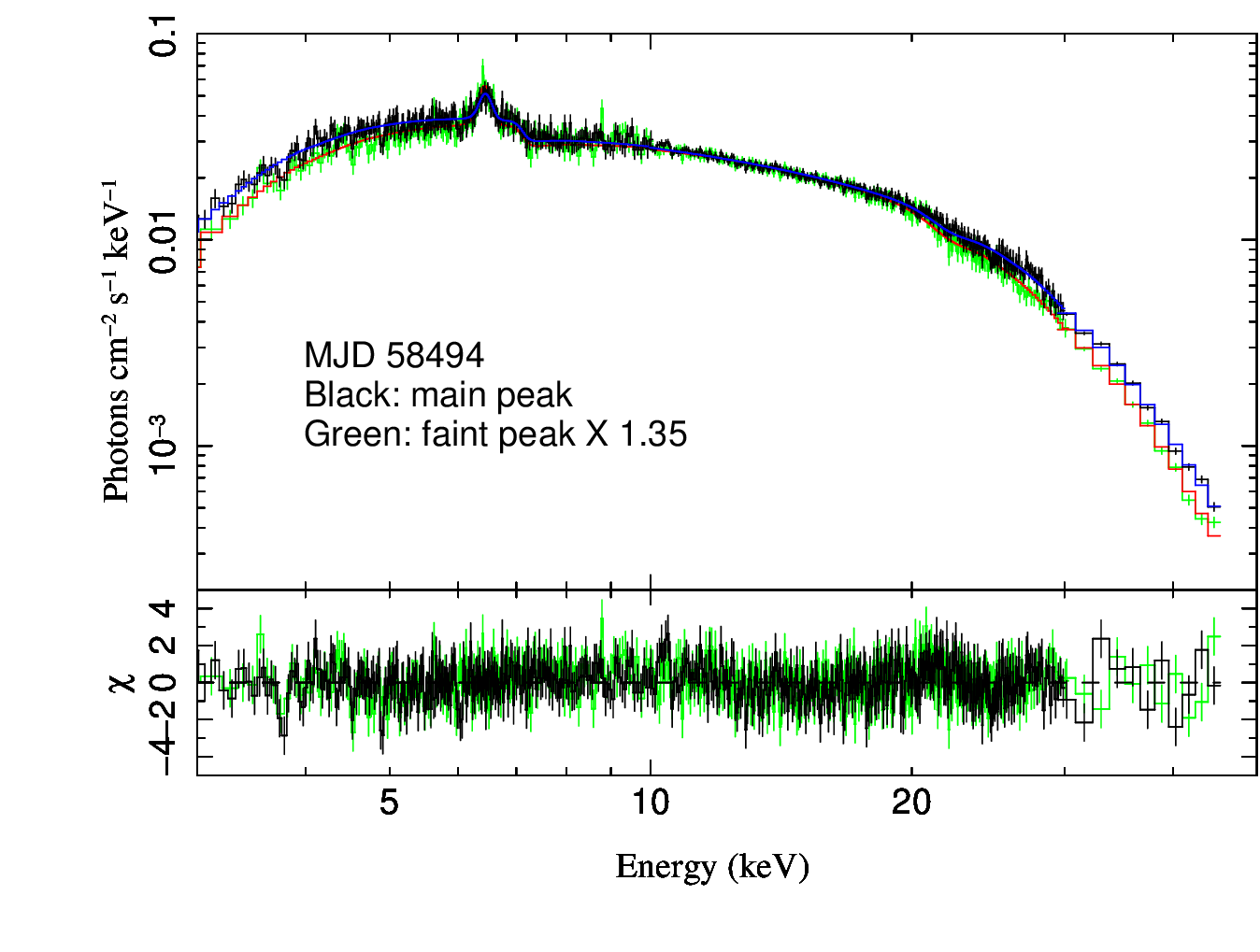}
	\hspace{-0.0cm}
	\includegraphics[width=3.4in]{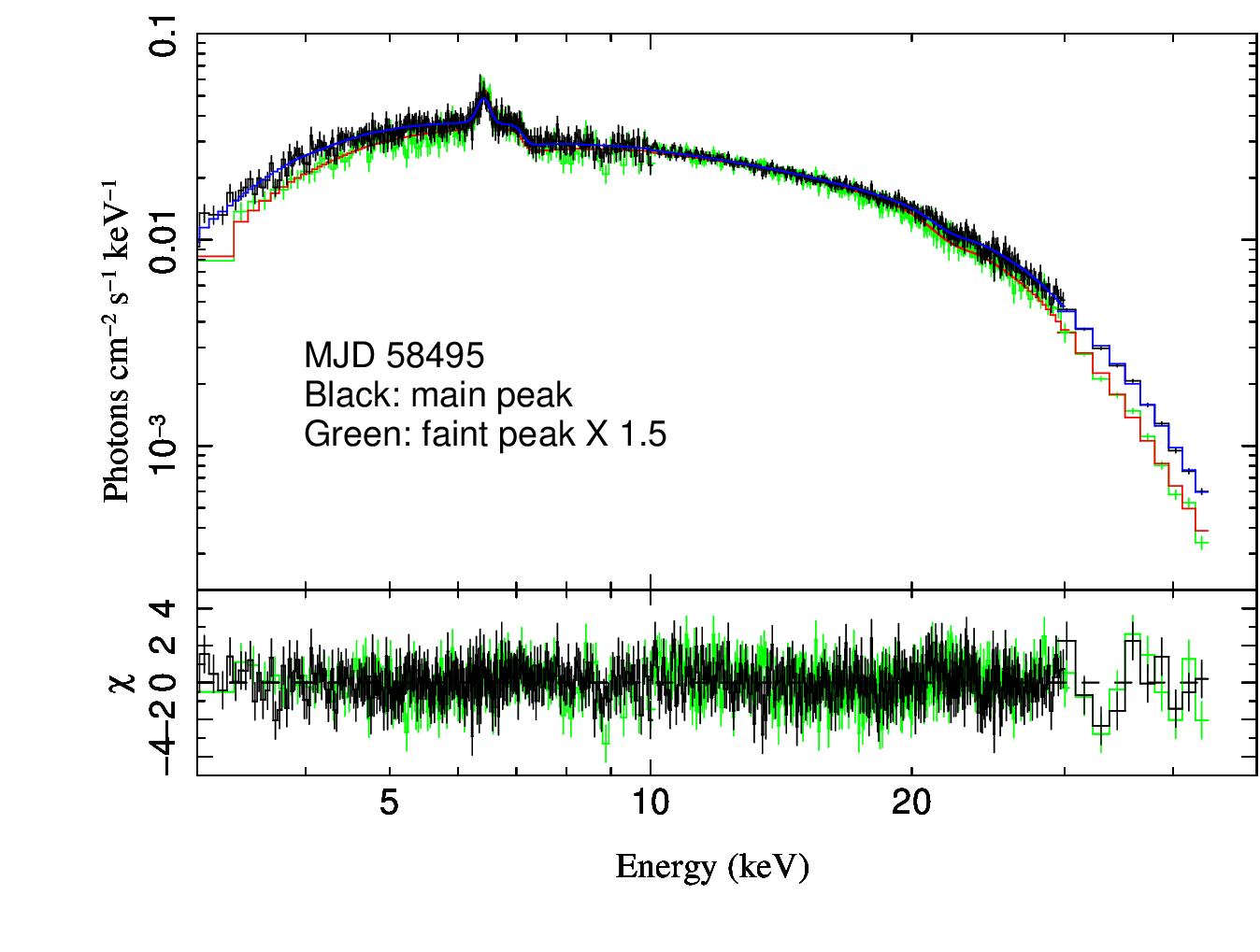}
	\includegraphics[width=3.4in]{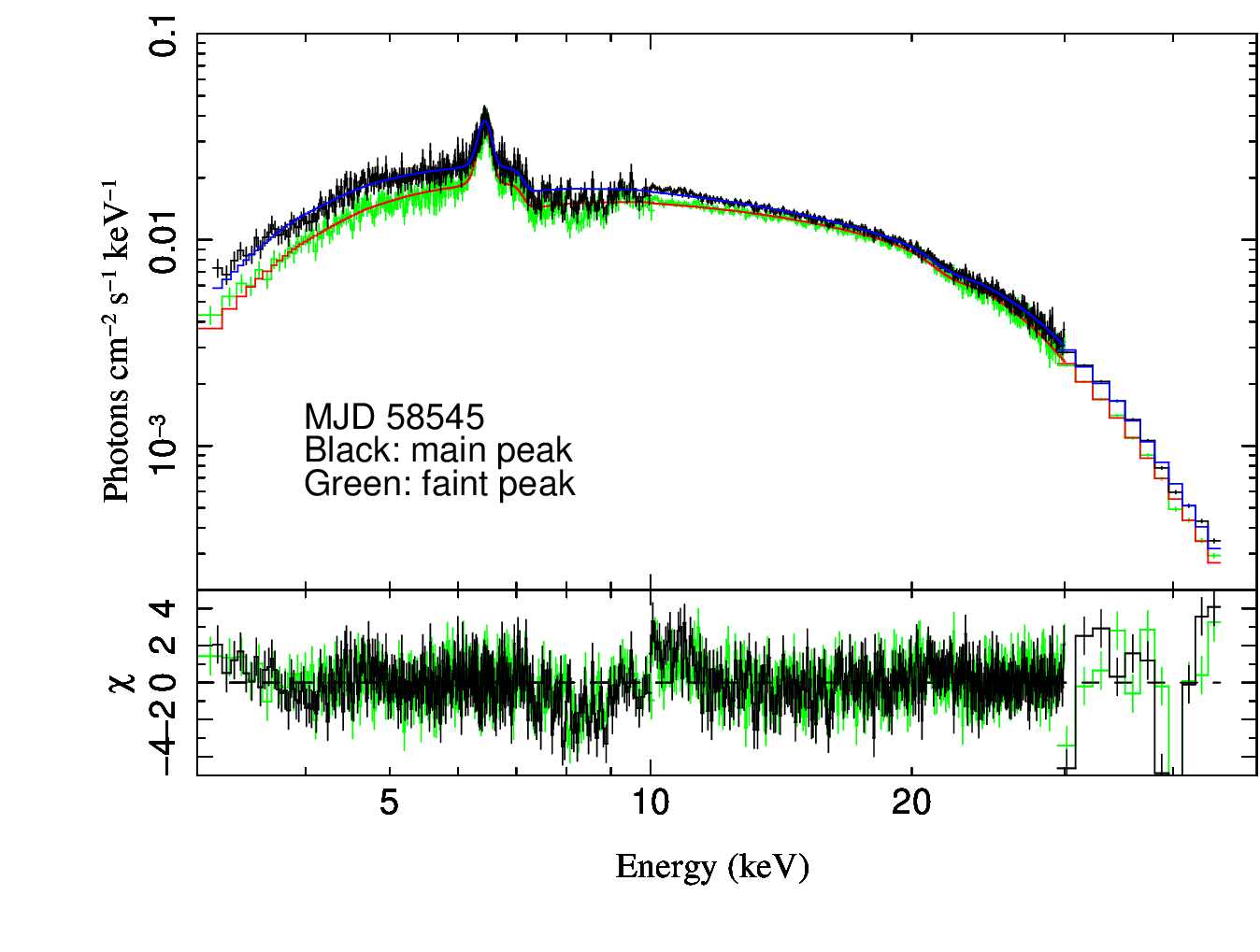}
	\includegraphics[width=3.4in]{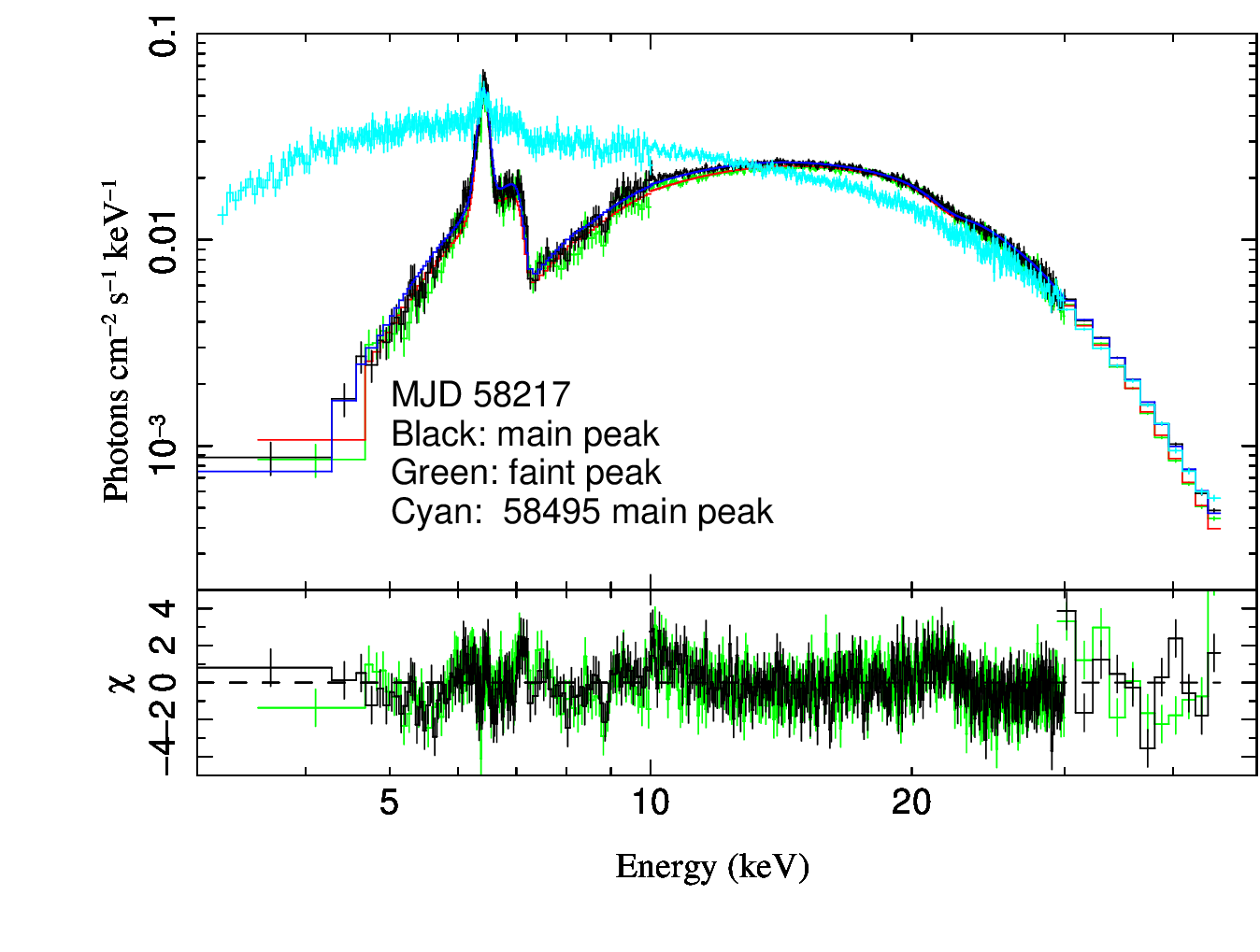}
	\caption{\hxmt broadband spectra of the main and faint peaks during the rapid spin-up 
	period MJD 58494 and 58495, the slow spin-up period MJD 58545, and 
	the wind-fed state MJD 58217. For comparison, the spectra of faint 
	peaks of MJD 58494 and 58495 are multiplied by a factor to make their 
	fluxes similar as the main peaks around 15 keV.
}
\end{figure*}

\begin{table*}
    \centering
	 \caption{Fitting results of the model of Eq. 1}
	 \tabcolsep 2.5pt
    \begin{tabular}{ c c c c c c c c c c c c}
        \hline
		 Peak & $N_H$ & A(PL) &$\Gamma({\rm PL})$ & E$_{cut}$ & E$_{fold}$ &  E$_{cyc}$&$\sigma_{cyc}$&D$_{cyc}$ &Norm(Fe) &$\chi^2_\nu (dof)$ & $L_{1-50 keV}$\\
		  & $10^{22}\,$cm$^{-2}$ & & & keV &  keV& & keV & keV & & & $10^{37}$erg/s\\
        \hline
		 58494 main & $15.1\pm0.7$ & $0.24\pm0.03$ & $0.83\pm0.05$ & $26.9\pm1.0$ & $5.9\pm0.3$ & $32.7\pm1.9$&$4.4\pm1.5$&$2.2\pm1.5$& $3.5\pm0.6$ &0.98 (548)& 4.4\\
		 58494 faint & $15.9\pm1.0$ & $0.14\pm0.02$ & $0.70\pm0.06$ & $23.6\pm1.0$ & $5.8\pm0.2$ & $30.8\pm3.4$&$3.8\pm2.3$&$1.3\pm1.3$&$4.1\pm0.5$ &0.99 (456)&2.4\\
		 58495 main & $15.3\pm0.7$ & $0.22\pm0.02$ & $0.79\pm0.04$ & $26.3\pm0.8$ & $5.8\pm0.2$ &  $32.1\pm2.0$&$3.3\pm2.2$&$0.8\pm0.8$&$3.2\pm0.6$ &0.92 (573)&4.1 \\
		 58495 faint & $17.6\pm1.3$ & $0.12\pm0.02$ & $0.72\pm0.07$ & $23.7\pm0.8$ & $5.4\pm0.1$ & $32.1$& 3.3 &$0.9\pm0.5$ &$3.4\pm0.5$ &1.10 (400)&2.2\\
		 58545 main & $18.5\pm0.7$ & $0.13\pm0.01$ & $0.77\pm0.05$ & $26.3\pm0.9$ & $5.9\pm0.1$ & 32 & $4.9\pm1.0$& $1.3\pm0.8$ &$4.5\pm0.3$ &1.41 (640)&2.7\\
		 58545 faint & $19.5\pm1.0$ & $0.09\pm0.01$ & $0.63\pm0.05$ & $25.4\pm1.0$ & $5.6\pm0.2$ &$32.0\pm0.9$ & $4.6\pm1.1$& $2.6\pm1.3$ & $4.7\pm0.3$ &1.16 (578)&2.2\\
		 58217 main & $87\pm5$ & $0.28\pm0.09$ & $0.72\pm0.12$ & $24.3\pm1.5$ & $5.5\pm0.2$ &  $32.5\pm1.0$ & $5.8\pm1.1$& $3.0\pm1.9$ &$9.4\pm0.4$ &1.28 (483)&5.1\\
		 58217 faint & $84\pm5$ & $0.17\pm0.05$ & $0.54\pm0.12$ & $23.6\pm1.1$ & $5.2\pm0.1$ &  $32.2\pm1.0$ & $5.8$& $3.0\pm0.6$ &$9.0\pm0.4$ &1.46 (468)&4.3\\
        \hline
 \end{tabular}
\begin{description}
\begin{footnotesize}
\item
Note: A(PL) is in units of $\rm photons\,keV^{-1}\,$cm$^{-2}\,s^{-1}$ at 1\,keV, 
	Norm(Fe) is in units of $10^{-3}\rm\,photons\,$cm$^{-2}\,s^{-1}$, and some CRSF energies
	and widths are not well constrained and they are fixed as those of the other peak.
\end{footnotesize}
\end{description}

\end{table*}

\section{Spectral analysis}

As illustrated in Figure 2, the pulse profiles during the spin-up episode
are energy dependent.
To study the energy spectrum of the main and faint peaks, 
we extracted the spectrum of the main and faint peaks from both MJD 58494 
and 58495 (D) observations, separately. They are plotted in the top panel of Figure 6.
To help the comparison, the spectra of faint peaks are multiplied by a factor to 
make the fluxes of faint peaks around 15 keV similar to those 
of main peaks. The spectra of main peaks and normalized faint peaks are 
quite similar, although show a slightly different curvature,
for both MJD 58494 and 58495.

To model the observed spectra, following \citet{Kre04}, we fit an absorbed power-law 
model with a Fermi-Dirac cutoff \citep[][]{Tan86}:
	 $1/\{\exp[(E-E_{cut})/E_{fold}]+1\}$,
where $E_{cut}$ and $E_{fold}$ are the cutoff and folding energy.
We add a Gaussian absorption line around 30-35 keV to model the CRSF.
We also add a Gaussian emission line around 6.4 keV to represent the Fe \Ka line, 
and its width is fixed as 0.01 keV.
The low energy absorption is modelled with tbabs \citep{Wil00}.
Thus, the adopted model is as follows:
\begin{equation}
	F(E)=FDcut*Gabs*tbabs*powerlaw+gauss.
\end{equation}
The fitting results are plotted in Figure 6 and listed in Table 2.
The absorption-corrected luminosities within 1-50 keV, assuming a distance 
of 3.5 kpc \citep{Nab19}, are also listed.

As can be seen, the model provides a reasonable fit to the observed spectra.
From Table 2 we see that the most apparent difference between the main and faint peaks
is the normalizations of the power-law model.
The differences of fitted cutoff energy and folding energy
between the main and faint peaks explain their spectral differences above 30 keV.
The absorption column densities of 
the faint peaks are a little larger than that of the main peak, which 
explain the relatively lower continua of the faint peaks at low
energy. We note that the differences of absorption between different 
peaks can not be due to variations far away from GX 301-2, but should originate 
in vicinity of the neutron star. 
While the inclusion of CRSF model improves the fitting results,
some CRSF parameters are not well constrained and are fixed as values of the other peak.

On the other hand, the fluxes of the Fe \Ka line are similar between
the main and faint peaks for both MJD 58494 and 58495. 
It implies that the illumination of the Fe K$\alpha$-emitting gas is similar during
the main and faint peaks. This could be obtained if the Fe K$\alpha$-emitting gas 
is quasi-symmetric with respect to the rotation direction of the neutron star. 

The spectra of the main and faint peaks of the slow spin-up period (MJD 58545, SD)
are plotted in the bottom left panel of Figure 6. They show similar behaviors as 
those of MJD 58494 and 58495 (D). The fitting residuals around 10 keV is large 
for MJD 58545 (SD). Such residuals could be due to instrument. 
Because different temperatures of ME detectors could affect the response of low energy
part and are not modeled in current version of data reduction pipeline, we have used LE
data for 8-10 keV. The 10-11 keV band of ME could still be affected.
On the other hand, the particle background of LE 
starts to become significant around 7-10 keV \citep{Liao20}. These two instrument effects 
could lead to the residual around 10 keV of MJD 58545 (SD), when the fluxes 
is lower than those of MJD 58494-58495 (D). 
Although with these caveats, the spectral shape of MJD 58545 (SD) is quite similar 
to those of MJD 58494 and 58495 (D).

The spectra of wind-fed state observation (MJD 58217, W) are plotted in the bottom right
panel of Figure 6. For comparison, the main peak spectrum of MJD 58495 (D)
is over-plotted. As can be seen, the spectra of the two peaks of
MJD 58217 are quite similar. They are also similar to that 
of the main peak of MJD 58495 above 30 keV. 
The 6.4 keV Fe \Ka line of MJD 58217 is very prominent, together 
with a sharp absorption edge around 7 keV. These spectral features of MJD 58217 are
typical of Compton reprocessed emission from neutral-like gas.
We also fit the wind-fed state spectra with the model of Eq.1, and the results 
are plotted with the observed spectra and listed in Table 2. The best-fitted
absorption column densities are around $8.5\times10^{23}$\,cm$^{-2}$, 
much larger than those of the spin-up period. The equivalent widths (EW) of 
the Fe \Ka line are
about 0.7 keV, also much larger than those of the spin-up period ($\sim0.15$ keV).
The residuals around 7 keV are the Fe \Kb line, which is not modeled in Eq. 1.
The residuals around 23 keV could be due to phase-dependent features 
of high energy cutoff \citep[e.g.][]{Fur18}, and here the phases are only divided into
main and faint peaks. The residuals around 10 keV could be due to 
instrument as discussed above. It is interesting to note that 10 keV bumps
have been observed on the spectrum of many X-ray pulsars 
\citep[e.g.][]{Cob02} and could be due to imperfect model of the
continuum of X-ray pulsars.

\section{Discussion and conclusion}

We studied the temporal and spectral differences of GX 301-2 between the spin-up
state and the wind-fed non spin-up state. During the initial 
rapid spin-up period of the 2019 spin-up event 
of GX 301-2, the main peak is higher than the faint peak.
During the late slow spin-up period, the pulse profiles are composed of two 
similar peaks, similar to those of wind-fed state. 
The line-of-sight absorption column densities are about $1.5\times10^{23}$cm$^{-2}$
for spin-up periods, and $\sim9\times10^{23}$cm$^{-2}$ for wind-fed state of 
similar fluxes.
The EWs of Fe \Ka line are about 0.15 keV for spin-up periods, and $\sim0.7$ keV 
for wind-fed state of similar fluxes.

For wind accretion, after passing through a bow shock, the wind material is supposed
to fall down
quasi-spherically to the magnetosphere of the neutron star \citep{DO73}. As a result, 
the line-of-sight direction traces a typical column density of the material 
around the neutron star. In contrast, if the material is accreted through 
a disk, most of the material is condensed into a disk, and the line-of-sight direction
could trace a much less column density if the disk plane is not in the line-of-sight
direction.
Since similar fluxes between the spin-up period and the wind-fed state indicate
similar mass accretion rates, the much less absorption column densities 
during the spin-up period support the scenario that the material is accreted through a 
disk-like structure, which spins up the neutron star.

The pulse profiles of GX 301-2 during the spin-up period show an interesting 
trend with luminosity: the main peak is similar to the faint peak 
during the slow spin-up period (with a luminosity $\sim2\times10^{37}$erg/s),
but the main peak increases with the observed fluxes during the rapid spin-up period
(with a luminosity $\sim4\times10^{37}$erg/s), while the faint peak keeps almost 
constant. Such a trend is confirmed by \fermi/GBM data at even higher 
luminosities ($\sim8\times10^{37}$erg/s). But the pulse profiles of wind-fed 
state show no such a trend. These behaviors could be related to the emitting 
region/pattern at different luminosities. 

For X-ray pulsars with low luminosity, the radiation is generally supposed to have a
pencil beam due to anisotropic scattering cross-section of photons in a strong magnetic 
field. While above certain luminosity, 
an accretion column is formed, and the radiation is 
supposed to have a fan beam. The pulse profiles of GX 301-2 always showed 
a two-peaked feature from the lowest to the highest observed luminosities 
\citep[e.g.][]{Lab05, Eva10}. The simplest appropriate model is 
to assume that each peak represents a pole, and each pole emits in 
a pencil beam \citep[e.g.][]{Wang81}.
The main peak corresponds to the pole close to the observer, while the faint
peak corresponds to the further pole. If both poles produce a similar 
pencil-like beam, the observed increasing trend of the main peak with 
increasing luminosity indicates that the emission along the polar direction 
is increasing, while the little change of the faint peak indicates that
the emission perpendicular to the polar direction is not changing much 
with increasing luminosity.

The pulse profiles of the neutron star could be related to luminosity 
in several different ways.
The magnetosphere radius of an accreting neutron star depends on the B-field 
and luminosity as follows \citep[e.g.][]{Fra02}:
\begin{equation}
	R_m=2.8\times10^8\Lambda B_{12}^{4/7}L_{37}^{-2/7} cm,
\end{equation}
where $\Lambda=1$ for wind accretion and $\Lambda<1$ for disk accretion, and 
we have adopted a neutron star mass of 1.5 $M_\odot$ and radius of 10 km.
The field line around the magnetosphere crosses the neutron star
surface with an angle $\beta$ relative to the magnetic axis:
\begin{equation}
	\sin^2\beta=\frac{R_{NS}}{R_m}\sin^2\alpha,
\end{equation}
where $\alpha$ is the angle between the disk plane and the magnetic axis.
Therefore, the higher the luminosity, the larger the $\beta$.
For wind accretion, the accreted material fills the whole column centered 
around the magnetic axis, and $\beta$ is the radius of the column;
while for disk accretion, the accreted material
is confined to a narrow wall, which is away from the magnetic axis with 
an angle $\beta$ \citep[e.g.][]{BS75,BS76}. 
The changing direction of the hot spot with luminosity for disk accretion
will affect the observed pulse profiles.

Another effect of different luminosities is radiation feedback.
For X-ray pulsars with high luminosity, the accreted gas flow is decelerated by 
the radiation pressure, and a radiative shock is formed, below which the material
is sinking down to the surface of neutron star \citep{BS76}. 
The critical luminosity depends on accretion process and the magnetic field, and is 
suggested to be around
$1\times10^{37}$ erg\,s$^{-1}$ for a cyclotron energy around 
30 keV, based on the bimodal variation of the cyclotron
energy with luminosity \citep[e.g.][]{Bec12, Mus15, Dor17}. Therefore, 
it is possible that the emission pattern changes with luminosity around the observed 
luminosities. 
Sudden transitions of pulse profiles have been observed around 
$10^{38}$ erg\,s$^{-1}$ for several sources \citep[e.g.][]{Wil18,Dor20}, and 
the exact value of critical luminosity is not well understood. 
We emphasize that feedback of radiation is expected to alter the geometry of 
the accretion disk/flow and such changes have indeed been observed 
\citep[e.g.][]{Wil18,Dor20}. Similar scenario could potentially apply 
also to GX 301-2. 

On the other hand, the trend of only main peak changing with luminosity is 
not observed for wind state at luminosities around $4-8\times10^{37}$ erg\,s$^{-1}$.
This could be due to the intrinsic differences between disk and wind accretion.
As mentioned above, the filled column of wind accretion always lies along the 
magnetic axis, and only the radius of the column changes with luminosity, therefore,
the inclination angle does not change with luminosity, different from disk accretion.
At similar luminosity, the area of the hot spot of wind accretion is 
generally larger than that of disk accretion, and thus, the critical 
luminosity is different for disk and wind accretion \citep[e.g.][]{Mus15}.
This could lead to different emission pattern for disk and wind accretion
at similar luminosity. 
We note that the above discussions assumed a simplified di-polar geometry
of the magnetic field, and in reality, other effects, such as 
asymmetric magnetic field and gravitational light bending, could also 
play a role \citep[e.g.][]{Fta86,Bil19,Iwa19}. Detailed calculations 
of the pulse profile shapes for X-ray pulsars are required to reveal whether 
these processes can explain 
the observed different trend of pulse profiles for disk and wind accretion.

\section*{Acknowledgements}
We thank our referee for his/her helpful suggestions.
This work made use of data from the {\it Insight}-\hxmt mission, a project 
funded by China National Space Administration (CNSA) and the Chinese 
Academy of Sciences (CAS), and also used data from \fermi/GBM and \swift/BAT.
This work is supported by the National Key R\&D
Program of China (2016YFA0400800) and the National Natural Science Foundation 
of China under grants U1938113, 11773035, U1838201, U1938101, U1838202, 11473027,
 11733009, U1838115, and is partially supported by the Scholar Program of Beijing Academy 
of Science and Technology (DZ BS202002).
VD acknowledges support by the Russian Science Foundation (grant 19-12-00423).

\section*{Data Availability}
The data underlying this article are publicly available at http://archive.hxmt.cn.
\bibliographystyle{mn2e}

\appendix

\end{document}